\documentclass[aps,pra,10pt,a4paper,twocolumn, superscriptaddress, floatfix]{revtex4-2}

\usepackage{graphicx} %
\usepackage{amsmath}
\usepackage{amsfonts}
\usepackage{amssymb}
\usepackage{physics}
\usepackage{tabularx}
\usepackage{booktabs}
\newcolumntype{L}{>{\raggedright\arraybackslash}X}

\usepackage[separate-uncertainty=true]{siunitx}
\DeclareSIUnit\sq{\ensuremath\Box}
\DeclareSIUnit{\sqrthz}{\ensuremath{\sqrt{\mathrm{\hertz}}}}
\DeclareSIUnit{\sample}{S}

\usepackage{hyperref}
\usepackage{xcolor}
\hypersetup{
    colorlinks,
    linkcolor={red!50!black},
    citecolor={blue!50!black},
    urlcolor={blue!80!black}
}

\newcommand{\kex}{\ensuremath{ \kappa_{\mathrm{ext}} }}
\newcommand{\Zex}{\ensuremath{ Z_{\mathrm{ext}} }}
\newcommand{\kB}{\ensuremath{ k_{\mathrm{B}} } }

\newcommand{\Lk}{\ensuremath{ L_{\mathrm{k}} } }
\newcommand{\London}{\ensuremath{ \lambda_{\mathrm{L}} } }

\date{June 14, 2024}

\begin{document}

\title{Temperature dependence of microwave losses in lumped-element resonators made from superconducting nanowires with high kinetic inductance}

\author{Ermes Scarano}
\affiliation{Department of Applied Physics, KTH Royal Institute of Technology, Hannes Alfvéns väg 12, SE-114 19 Stockholm, Sweden}
\author{Elisabet Arvidsson}
\affiliation{Department of Applied Physics, KTH Royal Institute of Technology, Hannes Alfvéns väg 12, SE-114 19 Stockholm, Sweden}
\author{August Roos}
\affiliation{Department of Applied Physics, KTH Royal Institute of Technology, Hannes Alfvéns väg 12, SE-114 19 Stockholm, Sweden}
\author{Erik Holmgren}
\affiliation{Department of Applied Physics, KTH Royal Institute of Technology, Hannes Alfvéns väg 12, SE-114 19 Stockholm, Sweden}
\author{David Haviland}
\email{haviland@kth.se}
\affiliation{Department of Applied Physics, KTH Royal Institute of Technology, Hannes Alfvéns väg 12, SE-114 19 Stockholm, Sweden}

\begin{abstract}

We study the response of several microwave resonators made from superconducting NbTiN thin-film meandering nanowires with large kinetic inductance, having different circuit topology and coupling to the transmission line. Reflection measurements reveal the parameters of the circuit and analysis of their temperature dependence in the range \SIrange{1.7}{6}{\kelvin} extract the superconducting energy gap and critical temperature. The lumped-element LC resonator, valid in our frequency range of interest, allows us to predict the quasiparticle contribution to internal loss, independent of circuit topology and characteristic impedance. Our analysis shows that the internal quality factor is limited not by thermal-equilibrium quasiparticles, but an additional temperature-dependent source of internal microwave loss.
    
\end{abstract}

\maketitle

\section{Introduction}

Superconducting nanowires with large kinetic inductance are interesting for applications in quantum technology due to their low loss and adjustable-strength nonlinearity. Their small size and low capacitance open the possibility of achieving high characteristic impedance, in some cases higher than the quantum resistance $R_{Q} = h/4e^{2}$ = \SI{6.45}{\kilo\ohm}~\cite{niepce2019nbn, grunhaupt2019granularaluminiumforquantumcircuits, shaikhaidarov2022quantized, rieger2023graluminium}. High kinetic inductance nanowires are fabricated from thin films with a single process step, resulting in compact, robust and reproducible resonant circuits in a microwave band, e.g. \SIrange{4}{8}{\giga\hertz}. As lumped elements, nanowire inductors provide a versatile platform for the design of quantum devices, sensors and detectors. Here we explore the properties of such lumped elements, with a particular focus on understanding microwave losses and the role of thermal-equilibrium quasiparticles (QP). 

Materials such as Nb~\cite{frunzio2005nb}, NbN~\cite{niepce2019nbn, anferov2020nbn, frasca2023nbn}, TiN~\cite{leduc2010tin, swenson2013tin, joshi2022tin}, and NbTiN~\cite{tanner2010nbtin, barends2010nbtin, hoeom2012nbtin, samkharadze2016nbtin, steinhauer2020nbtin, burdastyh2020nbtin, bretz-sullivan2022nbtin, parker2022nbtin} feature critical temperatures on the order of \SI{10}{\kelvin}, making them good candidates for applications in devices such as superconducting single-photon detectors (SSPDs)~\cite{korzh2020sspd}, Microwave Kinetic Inductance Detectors (MKIDs)~\cite{day2003mkid,zmuidzinas2012review}, Kinetic Inductance Parametric Amplifiers (KIPAs)~\cite{parker2022nbtin} and force transducers~\cite{roos2023kimec}. Such devices are routinely operated above \SI{1}{\kelvin} where microwave losses are influenced by thermal-equilibrium QPs , whose concentration increases exponentially with temperature. At lower temperatures losses are typically attributed to two-level systems (TLS) and out-of-equilibrium QPs.

Understanding microwave losses in high kinetic inductance materials requires consideration of the resonator design and coupling to the transmission line through which we measure its physical response. A lumped-element model allows us to relate values of the individual circuit variables to the scattering parameters of a reflection measurement. We study the temperature dependence of these variables in the range \SIrange{1.7}{6}{\kelvin}, fitting two models to the measured data: the traditional two-fluid model (TTF)~\cite{rammer1988magnetic} and the model based on the Mattis-Bardeen (MB) equations~\cite{bardeen1957theory}. The temperature dependence of the MB model accurately describes the behavior of NbTiN as a strong-coupling, disordered BCS superconductor, allowing us to determine the critical temperature $T_{c}$ and the energy gap $\Delta_{0}$. We analyze resonators with both series and parallel circuit topologies, and with different couplings to the microwave transmission line.

\begin{figure*}
    \centering
    \includegraphics[width=\linewidth]{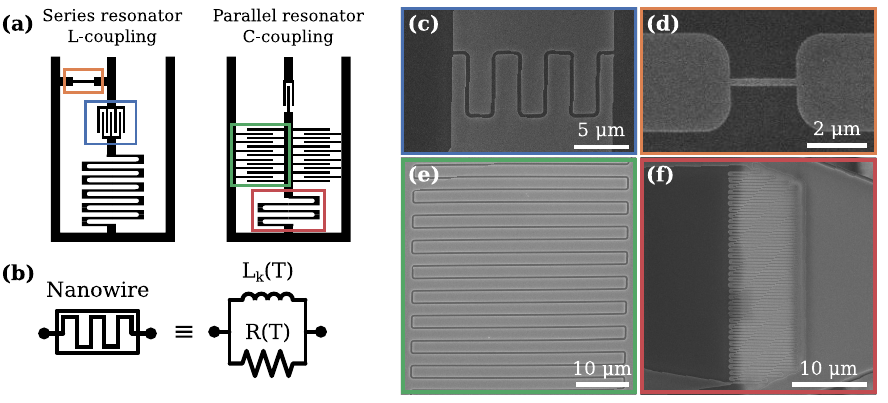}
    \caption{\label{fig:schematic_device_layout}
        \textbf{(a)} The device layout for a series resonator with inductive coupling and a parallel resonator with capacitive coupling. Scanning electron micrographs (SEM) of the components inside the colored boxes are shown in (c)-(f).
        \textbf{(b)}  Lumped-element model of the meandering nanowire kinetic inductor consisting of two parallel channels: a temperature-dependent inductance $\Lk(T)$, and a temperature-dependent resistance $R(T)$.
        \textbf{(c)} An SEM image of the interdigital series capacitor.
        \textbf{(d)} An SEM image of the short nanowire shunting inductor, for inductive coupling of the series resonator.
        \textbf{(e)} An SEM image of an interdigital parallel capacitor for the parallel resonator.
        \textbf{(f)} An SEM image of a meandering nanowire kinetic inductor used in both designs.
    }
\end{figure*}

Combining the lumped-element circuit description and two-fluid model, we are able to unambiguously quantify the contribution to microwave loss. As expected for an accurate lumped-element model, our analysis shows that the QP loss, at a given resonance frequency, is independent of the circuit topology and the values of the lumped elements. Comparing the data to the prediction of the model, we show that our resonators contain an additional temperature-dependent loss contribution which is not explained by thermal-equilibrium QPs.

\section{Devices and models}

We realize high-Q factor resonators in both series and parallel circuit topologies by combining high kinetic inductance superconducting nanowires and interdigital capacitors. Figure~\ref{fig:schematic_device_layout}(a) shows a illustrative layout of the two circuit topologies and Fig.~\ref{fig:schematic_device_layout}(b) the lumped-element model of the kinetic inductor. Figure~\ref{fig:schematic_device_layout}(c)-(f) show scanning electron micrographs of the individual lumped elements. The circuits are fabricated from a single layer of NbTiN film with thickness $t$ = \SI{15}{\nano\metre} on a \SI{600}{\nano\metre}-thick substrate of silicon nitride (SiN) on silicon (Si). The superconducting film was co-sputtered with argon (Ar) and nitrogen (N$_2$) gas from separate Nb and Ti targets. The patterns are defined with a combination of photolithography and e-beam lithography and realized with reactive ion etching. Details of design and fabrication of these types of devices can be found in \cite{roos2024design}.  The resonators are either directly coupled to a transmission line or coupled via a reactive element (capacitor or inductor, depending on topology) to adjust the total quality factor and coupling coefficient $\eta = \kex / (\kex + \kappa_{0})$, where $\kex$ is the loss rate to the external transmission line, and $\kappa_{0}$ the loss rate internal to the resonator.

Our description of the kinetic inductor as a single lumped-element is valid for frequencies much lower than the first eigenmode of the meandering nanowire~\cite{roos2024design}, $\sim \SI{24}{\giga\hertz}$ for the design shown in Fig.~\ref{fig:schematic_device_layout}(f). Disordered superconductors such as NbTiN have low Cooper pair density $n_s$  and  large London penetration depth $\London \propto 1 / \sqrt{n_{s}} \approx$~\SI{250}{\nano\metre} \cite{khan2022characterization,watanabe1994kinetic}. In the regime of small thickness $t \ll \London$ the nanowire's kinetic inductance per unit length far exceeds its geometric inductance. 

The measurement setup is described in the Appendix. We measure the reflection coefficient,
\begin{equation}
    \Gamma(\omega) = \frac{\Zex - Z_{R}(\omega)}{\Zex + Z_{R}(\omega)},
    \label{eqn:gamma_impedances}
\end{equation}
where $Z_{R}(\omega)$ is impedance of the resonator and $\Zex$ is the impedance seen by the resonator. For a single port high-Q resonator, $\Gamma(\omega)$ can be expressed as \cite{pozar2011microwave,rieger2023microwaveresonatorfano}
\begin{equation}
    \Gamma(\omega) = \frac{ \kappa_{0} - \kex + i 2 (\omega - \omega_{0}) }{ \kappa_{0} + \kex + i 2 (\omega - \omega_{0})},
    \label{eqn:reflection_coefficient}
\end{equation}
where the resonance frequency $\omega_{0}$, internal loss rate $\kappa_{0}$ and external loss rate $\kex$, are the key figures-of-merit that we use to compare different resonators. 

\begin{figure*}
    \centering
    \includegraphics[width=0.85\linewidth]{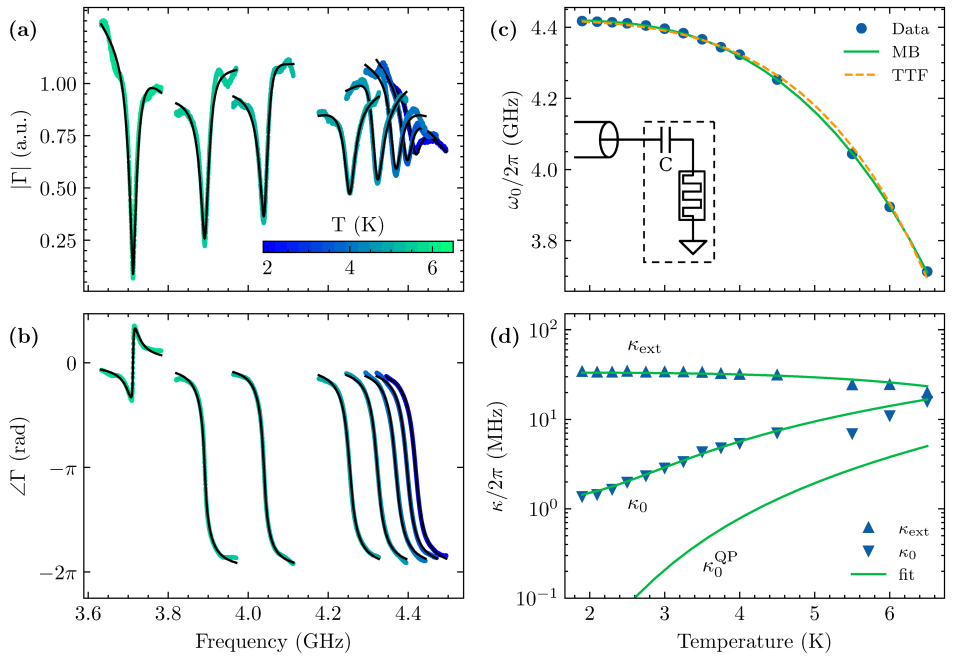}
    \caption{\label{fig:series_direct_coupling}
        \textbf{(a)} Magnitude and \textbf{(b)} phase of the reflection coefficient $\Gamma(\omega)$ of a directly coupled series $LC$-resonator measured at multiple temperatures with an input power of -102 dBm. Black solid lines are best fit curves to the data.\textbf{(c)} Temperature dependence of the resonance frequency $\omega_{0}$  with fits using the Mattis-Bardeen (MB) model and the traditional two-fluid model (TTF). The inset displays the equivalent circuit diagram.
        \textbf{(d)} External $\kex$ and internal $\kappa_{0}$ decay rates with fits using the MB model. The lowest solid green line traces the predicted internal losses due to the QP contribution in the two-fluid model $\kappa_{0}^{\mathrm{QP}}$.
    }
\end{figure*}

The lumped-element model allows us to relate the resonator parameters to individual circuit variables. For a series topology, we have~\cite{pozar2011microwave},
\begin{equation}
    \omega_{0} = \frac{1}{\sqrt{LC}},~
    \kex = \frac{\Re[\Zex]}{L},~
    \kappa_{0} =  \frac{R_{s}}{L},
    \label{eqn:series_topology}
\end{equation}
and for parallel we have
\begin{equation}
    \omega_{0} = \frac{1}{\sqrt{LC}},~
    \kex = \frac{1}{\Re[\Zex] C},~
    \kappa_{0}  = \frac{1}{R_{p}C},
    \label{eqn:parallel_topology}
\end{equation}
where $R_{s}$ ($R_{p}$) is an effective resistance describing losses. Losses in the lumped-element nanowire kinetic inductor are well modeled by a temperature-dependent parallel resistance $R(T)$, as shown in Fig.~\ref{fig:schematic_device_layout}(b). Thus $R_{p} = R(T)$ while $R_{s} = \Re \left[ (R(T)^{-1} + (i \omega \Lk)^{-1})^{-1} \right] = \omega_{0}^{2} \Lk^{2}/R(T)$.

\section{Two-fluid model}

Quasiparticle losses in the lumped-element kinetic inductor can be described by the two-fluid model of superconductivity~\cite{tinkham2004introduction}. This model invokes a complex conductivity, with a real part describing dissipation due to QP current, and an imaginary part describing the inertia of the condensate. The QP resistance is inversely proportional to the normal carrier density $R_{\mathrm{QP}} \propto 1/n_{n}$, and the kinetic inductance inversely proportional to the density of pairs, $\Lk \propto 1/n_{s}$. 

The concentration of both normal and superconducting carriers changes with temperature, but the total density of charge carriers $n = n_{s}(T) + n_{n}(T)$ is independent of temperature. With $n = n_{s}(0) = n_{n}(T_{c})$ we can explicitly write the temperature scaling of $\Lk$ and $R_{\mathrm{QP}}$,
\begin{equation}
    \frac{\Lk(T)}{\Lk(0)} = \frac{n}{n_{s}(T)} \equiv \Lambda(T),
    \label{eqn:lambda_temperature}
\end{equation}
and
\begin{equation}
    \frac{R_{\mathrm{QP}}(T)}{R_{\mathrm{N}}} = \frac{n}{n - n_{s}(T)} = \frac{\Lambda(T)}{\Lambda(T)-1},
    \label{eqn:resistance_quasiparticle}
\end{equation}
where $R_{\mathrm{N}} = R_{\mathrm{QP}}(T_{c})$ is the normal state resistance of the kinetic inductor.

We fit both the TTF and MB models to the measured data. The former predicts a quadratic dependence of the London penetration depth on temperature, leading to a power-four scaling of the kinetic inductance~\cite{tinkham2004introduction},
\begin{equation}
    \Lambda_{\mathrm{TTF}}(T) = \frac{1}{1-(T/T_{c})^4},
    \label{eqn:two-fluid_lambda}
\end{equation}
and latter relates the complex conductivity to the BCS energy gap. In the low frequency limit $\hbar \omega \ll \Delta$, where pair breaking from the driving field is negligible, the MB model predicts~\cite{annunziata2010nanoinductors},
\begin{equation}
    \Lambda_{\mathrm{MB}}(T)=\frac{\Delta(0)}{\Delta(T)} \coth \left( \frac{\Delta(T)}{2 \kB T} \right).
    \label{eqn:mattis-bardeen_lambda}
\end{equation}
In the fitting procedure we use an expression which is a good approximation to the temperature dependence of the superconducting energy gap of BCS superconductors~\cite{sheahen1966rulesforenergygap}:
\begin{equation}
    \frac{\Delta(T)}{\Delta(0)} = \sqrt{\cos\left(\frac{\pi}{2}\left(\frac{T}{T_{c}}\right)^{2}\right)}.
    \label{eqn:semi-empirical_delta}
\end{equation}

\begin{figure*}
    \centering
    \includegraphics[width=0.85\linewidth]{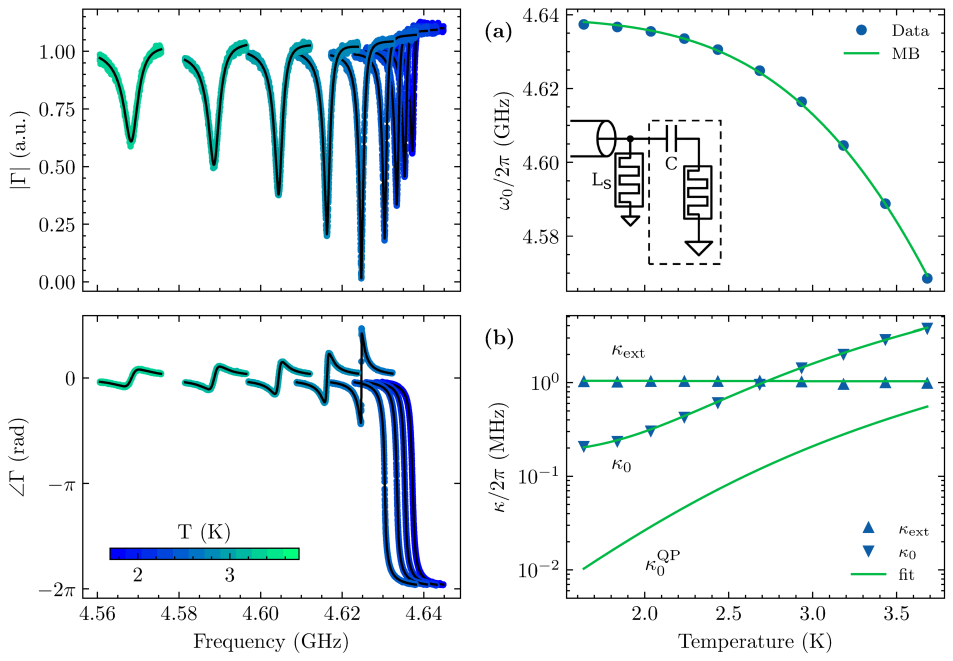}
    \caption{\label{fig:shunted_series}
        \textbf{(a)} Magnitude and \textbf{(b)} phase of the reflection coefficient $\Gamma(\omega)$ of an inductively coupled series $LC$-resonator, measured at multiple temperatures  with an input power of -122 dBm. Black solid lines are best fit curves to the data.\textbf{(c)} Temperature dependence of the resonance frequency $\omega_{0}$  with a fit using the Mattis-Bardeen (MB) model. The inset displays the equivalent circuit diagram.
        \textbf{(d)} External $\kex$ and internal $\kappa_{0}$ decay rates with fits using the MB model. The lowest solid green line traces the predicted internal losses due to the QP contribution in the two-fluid model $\kappa_{0}^{\mathrm{QP}}$.   
    }
\end{figure*}

\section{Data analysis}

Figure~\ref{fig:series_direct_coupling} shows the temperature dependence of the parameters of a series resonator obtained by fitting Eqn.~\eqref{eqn:reflection_coefficient} to the measured $\Gamma(\omega)$ at each temperature [Fig.~\ref{fig:series_direct_coupling}(a, b)]. The resonator consists of a single interdigital capacitor in series with a \SI{100}{\nano\metre}-wide nanowire kinetic inductor, directly coupled to the transmission line [see inset of Fig.~\ref{fig:series_direct_coupling}(c)]. We take the capacitor to be independent of temperature. For the directly coupled resonator we can analytically relate the resonator parameters to the lumped-element circuit variables, allowing us to express the temperature dependence of the resonance frequency
\begin{equation}
    \omega_{0}(T) = \frac{\omega_{0}(0) }{\sqrt{\Lambda(T)}}.
    \label{eqn:omega_0_temperature}
\end{equation}
We use Eqn.~\eqref{eqn:omega_0_temperature} to fit the TTF and MB models with $\Lambda(T)$ given by Eqn.~\eqref{eqn:two-fluid_lambda} and Eqn.~\eqref{eqn:mattis-bardeen_lambda}, respectively. The fits are shown in Fig.~\ref{fig:series_direct_coupling}(a) where the MB model shows excellent agreement to the data, with critical temperature $T_{c}$ = \SI{9.45}{\kelvin} and energy gap ratio $\Delta_{0} /\kB T_{c} = 1.85$. From this fit we conclude that NbTiN behaves as a strong coupling ``dirty'' BCS superconductor~\cite{linden1994modified}.

The temperature dependence of the external loss rate $\kex(T)$ is well explained with the same $\Lambda_{\mathrm{MB}}(T)$ found in the fit of $\omega_{0}(T)$ by adjusting only one scaling factor $\kex(0)$ [see Fig.~\ref{fig:series_direct_coupling}(d)],
\begin{equation}
    \kex(T) = \frac{\Re[\Zex]}{\Lk(0)\Lambda(T)} = \frac{\kex(0)}{\Lambda_{\mathrm{MB}}(T)}.
    \label{eqn:kappa_ext_temperature_dependence}
\end{equation}
Here we point out that $\kex \approx \SI{30}{\mega\hertz} \gg \kappa_{0}$ over most of the temperature range studied. The low quality factor of this over-coupled resonator makes the determination of $\Gamma(\omega)$ sensitive to details of $\Zex$. For example, standing waves from small impedance mismatches in the transmission line give a frequency dependent $\Zex(\omega)$ which may vary significantly over the bandwidth of the resonator, increasing the uncertainty in the determination of the circuit variables. 

To explain the temperature dependence of the internal loss rate $\kappa_{0}(T)$ we first consider only the QP contribution, i.e. $R = R_{\mathrm{QP}}(T)$. The series topology gives
\begin{equation}
    \kappa_{0}^{\mathrm{QP}} = \frac{R_{s}}{\Lk}=\frac{\omega_{0}^{2}(T)\Lk(T)}{R_{\mathrm{QP}}(T)}.
    \label{eqn:kappa_qp}
\end{equation}
One arrives at the same expression for the parallel topology, where $\kappa_{0}^{\mathrm{QP}}= (R_{\mathrm{QP}}(T)C)^{-1}$. Combining Eqn.~\eqref{eqn:lambda_temperature}, Eqn.~\eqref{eqn:resistance_quasiparticle} and Eqn.~\eqref{eqn:kappa_qp} with the relation between the zero-temperature kinetic inductance and the normal-state resistance~\cite{tinkham2004introduction}
\begin{equation}
    \Delta_{0} = \frac{\hbar R_{\mathrm{N}}}{\pi \Lk},
    \label{eqn:delta_0_normal_state_resistance}
\end{equation}
we arrive at an expression for the internal loss rate due to QPs,
\begin{equation}
    \kappa_{0}^{\mathrm{QP}} = \frac{\omega_{0}^2(0)\hbar}{\pi\Delta_{0}}\left(\frac{\Lambda(T)-1}{\Lambda(T)}\right).
    \label{eqn:kappa_qp_temperature}
\end{equation}
Eqn.~\eqref{eqn:kappa_qp_temperature} tells us that, for a given resonance frequency, the QP loss rate is independent of the resonator's characteristic impedance $Z_{c} = \sqrt{\Lk / C}$ and circuit topology.

We plot Eqn.~\eqref{eqn:kappa_qp_temperature} in Fig.~\ref{fig:series_direct_coupling}(d), where we see that the measured $\kappa_{0}$ is nearly an order of magnitude larger than $\kappa_{0}^\mathrm{QP}$. This implies that there are additional sources of microwave dissipation, e.g. radiative losses (not into the transmission line) or temperature-dependent dielectric losses in the insulating layer~\cite{gerhold1998properties}. We model these as additional resistive channels in parallel with the kinetic inductor, 
\begin{equation}
    \frac{1}{R(T)} = \frac{1}{R_{\mathrm{QP}}(T)} + \frac{1}{R_{d0}}+\frac{1}{R_{d} e^{\beta/T}}.
    \label{eqn:resistance_temperature}
\end{equation}
The parameter $R_{d0}$ describes residual losses at zero temperature, and the parameters $R_{d}$ and $\beta$ describe a thermally activated loss mechanism. Figure~\ref{fig:series_direct_coupling}(b) shows a best fit with the values of all variables given in Table~\ref{tab:circuit_parameters}.

For the case of a series resonator directly connected to a transmission line, the value of $\Lk = \kappa_{0}(0)/ \Re \left[ \Zex \right]$ in Table~\ref{tab:circuit_parameters} is found by assuming $\Re \left[ \Zex \right]$ = \SI{50}{\ohm}. As previously mentioned, this assumption is not very reliable and furthermore the strongly over-coupled resonator leads to intrinsic uncertainty in the determination of $\kappa_{0}(T)$~\cite{rieger2023microwaveresonatorfano}. More accurate analysis can be made closer to critical coupling where $\kex \approx \kappa_{0}$. Critical coupling is difficult to achieve with direct connection because typically $\Re \left[ Z_R(\omega_0) \right] \ll \SI{50}{\ohm}$. Adding a reactive circuit element allows us to tune the external impedance closer to critical coupling while increasing total quality factor.

\begin{figure*}
    \centering
    \includegraphics[width=0.85\linewidth]{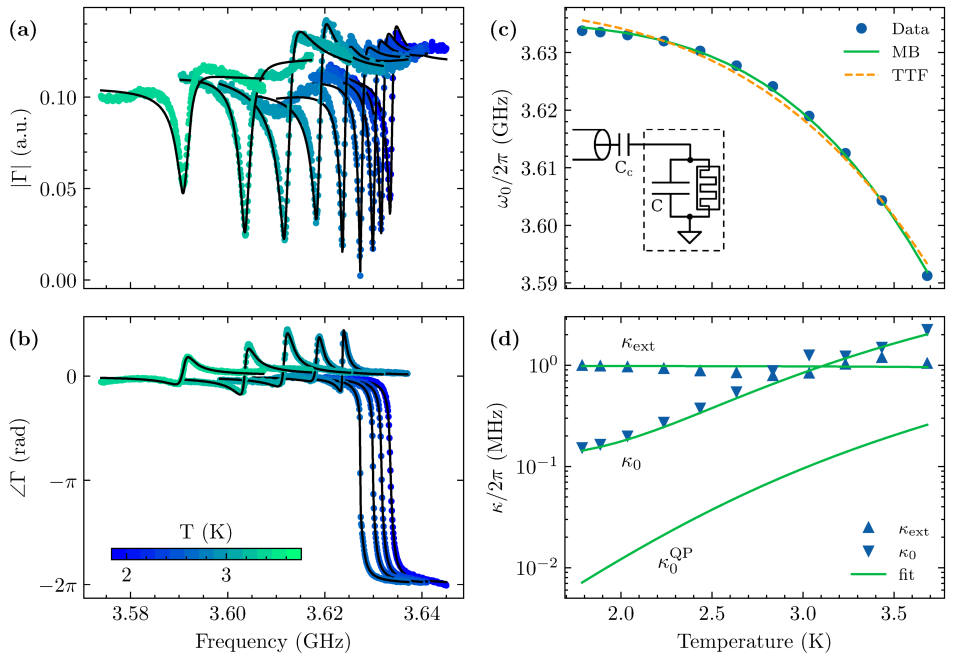}
    \caption{\label{fig:parallel}
        \textbf{(a)} Magnitude and \textbf{(b)} phase of the reflection coefficient $\Gamma(\omega)$ of a parallel $LC$-resonator measured at multiple temperatures  with an input power of -99.5 dBm. Black solid lines are best fit curves to the data.\textbf{(c)} Temperature dependence of the resonance frequency $\omega_{0}$ with fits using the Mattis-Bardeen (MB) model and the traditional two-fluid model (TTF). The inset displays the equivalent circuit diagram.
        \textbf{(d)} External $\kex$ and internal $\kappa_{0}$ decay rates with fits using the MB model. The lowest solid green line traces the predicted internal losses due to the QP contribution in the two-fluid model $\kappa_{0}^{\mathrm{QP}}$.
    }
\end{figure*}

\subsection*{Reactively coupled resonators}

The series resonator is reactively coupled to the transmission line with an inductive shunt $L_{s}$ to ground, realized by the short section of nanowire in the illustrative layout of Fig.~\ref{fig:schematic_device_layout}(a), and shown schematically in the inset of Fig.~\ref{fig:shunted_series}(c). This circuit is the dual of the more common capacitively-coupled parallel resonator, which we discuss later. The shunt inductance $L_{s}$ modifies the external load impedance as
\begin{eqnarray}
 \Zex &=& \left(\frac{1}{(\SI{50}{\ohm})} + \frac{1}{i\omega L_{s}}\right)^{-1}=\nonumber\\ &=&\frac{\omega^{2}L_{s}^{2} (\SI{50}{\ohm}) + i \omega L_{s} (\SI{50}{\ohm})^{2}}{(\SI{50}{\ohm})^{2}+\omega^{2}L_{s}^{2}}\approx \frac{\omega^{2}L_{s}^{2}}{(\SI{50}{\ohm})} + i \omega L_{s},
    \label{eqn:external_load_impedance}
\end{eqnarray}
with the approximation valid when $\omega L_{s} \ll \SI{50}{\ohm}$, which is the case for the range of frequency and temperature studied here. With $L_{s} \ll \Lk$ the resonator parameters become
\begin{eqnarray}
    \omega_{0} &=& \frac{1}{\sqrt{C(\Lk+L_{s})}},\\
    \kex &=& \frac{\omega_{0}^{2} L_{s}^{2}}{ (\SI{50}{\ohm}) (\Lk+L_{s})},\label{eqn:kexs}\\
    \kappa_{0} &=& \frac{\omega^2 \Lk^2 }{R(\Lk+L_{s})}.
\end{eqnarray}
where $R$, $\Lk$ and $L_{s}$ are temperature-dependent. Assuming $L_{s}$ and $\Lk$ have the same temperature scaling, Eqn.~\eqref{eqn:kexs} would predict $\kex$ to be independent of temperature.

The above expressions capture the qualitative behavior with temperature, but they are not accurate enough for fitting. Rather, we fit Eqn.~\eqref{eqn:gamma_impedances} to the measured $\Gamma(\omega)$ at all temperatures to obtain the circuit variables, without the approximations and assumptions stated above. Here we fit only the MB model, as the fit does not converge with the TTF model. The results are given in Table~\ref{tab:circuit_parameters} and the corresponding resonator parameters for this fit are shown by the solid lines in Fig.~\ref{fig:shunted_series}. As expected the fit shows a nearly constant $\kex \approx \SI{1}{\mega\hertz}$. In the range of temperature studied $\kappa_{0}(T)$ crosses $\kex$, with the resonator going from over-coupled at low temperature to under-coupled at higher temperature. The analysis spans a range of coupling coefficient $0.25 < \eta < 0.9$, where determination of the internal losses from the measured scattering parameters is most accurate. As for the directly coupled series resonator, we find the theoretically predicted $\kappa_{0}^{\mathrm{QP}}$ to be far lower than that determined from the measured scattering parameters.

We also studied the capacitively-coupled parallel $LC$-resonator, with a layout illustrated in Fig.~\ref{fig:schematic_device_layout}(a) and shown schematically in the inset of Fig.~\ref{fig:parallel}(c). The coupling capacitance $C_{c}$ transforms the external load admittance as
\begin{equation}
    Y_{\mathrm{ext}} = \frac{1}{\Zex} = \frac{1}{(\SI{50}{\ohm})+1/i\omega C_{c}}=\frac{\omega^{2}C_{c}^{2} (\SI{50}{\ohm}) + i\omega C_{c}}{\omega^{2}C_{c}^{2} (\SI{50}{\ohm})^{2} + 1}.
    \label{eqn:external_load_admittance}
\end{equation}
A high-Q factor requires a coupling capacitance such that near resonance $\omega C_{c} \ll 1 / \SI{50}{\ohm}$. In this case, the external admittance becomes
\begin{equation}
    Y_{\mathrm{ext}} = \omega^{2} C_{c}^{2} 50 + i \omega C_{c}.
    \label{eqn:external_load_admittance_simplified}
\end{equation}
The real part reduces the external decay rate, increasing the quality factor, while the imaginary part adds to the resonator capacitance. For $C_{c} \ll C$ there is a slight shift of the resonance frequency. The resonator parameters become
\begin{eqnarray}
    \omega_{0} &=& \frac{1}{\sqrt{\Lk(C+C_{c})}},\\
    \kex &=& \frac{\omega_{0}^{2} C_{c}^{2} (\SI{50}{\ohm})}{C+C_{c}},\\
    \kappa_{0} &=& \frac{1}{R(C+C_{c})}.
    \label{eqn:resonator_parameters_parallel}
\end{eqnarray}
with $\Lk$ and $R$ depending on temperature.

\begin{table}
    \centering
    \setlength\tabcolsep{0.75\tabcolsep}%
    \begin{tabularx}{\linewidth}{L|L L L L L L L L}
        \toprule
        & \textit{\Lk} (\si{\nano\henry}) & \textit{C} (\si{\femto\farad}) & $L_{s}$ (\si{\nano\henry}) & $C_{c}$ (\si{\femto\farad}) & $R_{\mathrm{N}}$ (\si{\kilo\ohm}) & $R_{d}$ (\si{\kilo\ohm}) & $R_{d0}$ (\si{\mega\ohm}) & $\beta$ (\si{\kelvin}) \\
        \midrule
        Fig.~\ref{fig:series_direct_coupling}$^{\dagger}$ & 238 &  5.44 & --- & --- & 1720 & 514 & 23 & 11\\
        Fig.~\ref{fig:shunted_series}   & 107  & 10.9 & 0.205 & --- & 786 & 805  & 83 & 15 \\
        Fig.~\ref{fig:parallel} & 2.38 & 805  & --- & 13.8 & 5.85  & 1.73 & 1.7  & 16 \\
        \bottomrule
    \end{tabularx}
    \caption{\label{tab:circuit_parameters}
        Best-fit parameters for the solid lines shown in the respective figures. $^{\dagger}$The directly coupled series resonator of Fig.~\ref{fig:series_direct_coupling} had low quality factor and was strongly over-coupled, leading to larger uncertainty.  
    }
\end{table}

For the capacitively-coupled parallel $LC$-resonator, four circuit variables define the three resonator parameters. The fit requires prior knowledge of at least one circuit variable. The data in Fig.~\ref{fig:parallel}(c) shows the best fit of $\omega_{0}(T)$ using the TFF and MB models. Again the MB model fits better than the TFF model, and we find that the measured internal loss rate $\kappa_{0}(T)$ exceeds the expected QP contribution by nearly an order of magnitude, as shown in Fig.~\ref{fig:parallel}(d). We also show the best fit of $\kappa_{0}(T)$ to the loss model for $R(T)$ given in Eqn.~\eqref{eqn:resistance_temperature} with best-fit values given in Table~\ref{tab:circuit_parameters}. Here we note that the residual resistance $R_{d0}$ for the parallel resonator is almost three orders of magnitude smaller that that for the series configuration, consistent with an increased contribution from dielectric losses due to the much larger resonator capacitance $C$.

\section{Conclusion}

In a temperature range where thermal-equilibrium quasiparticle (QP) contributions to the internal losses cannot be neglected, we analyzed the resonance frequency and loss rates of microwave resonators made from superconducting meandering nanowires with high kinetic inductance. We showed how microwave measurements of the resonator reflection coefficient in a relatively narrow temperature interval \SIrange{1.7}{6}{\kelvin} provide sufficient information to determine the critical temperature and superconducting energy gap. Quantifying these parameters is essential as they, for a given resonance frequency, uniquely determine the QP contribution to losses and therefore the maximum achievable quality factor at any given temperature, irrespective of the resonators characteristic impedance and circuit topology. 

We studied three lumped-element circuit topologies with different coupling to the transmission line. For each we show that the Mattis-Bardeen model provides a consistently good description of the temperature dependence of the kinetic inductance derived from the shift in resonance frequency, with no topology-specific additional assumption. Combined with a proper characterization of the measurement setup, our analysis allows us to quantify the zero-temperature kinetic inductance and normal-state resistance of the nanowire.

Our analysis showed the presence of additional internal microwave losses, not explained by thermal-equilibrium QPs. A comparison between the measured and predicted QP losses for different circuit topologies provides information about the origin of these additional losses. Compared to the series case, the reduced residual resistance of a parallel topology points to the presence of substrate dielectric losses, consistent with the much larger area of the interdigital capacitor.

Our study provides tools for those interested in circuit design with high kinetic inductance, including alternative approaches to achieve a desired coupling to the transmission line. From our analysis we conclude that there is significant room for increasing the quality factor of microwave resonators realised with high kinetic inductance nanowires. Future work should analyze nanowires of different materials, on different substrates, or suspended in vacuum. We hope that the methods and analysis presented here will inspire such future studies.

\section*{Data Availability Statement}
All raw and processed data, as well as supporting code for data processing and figure generation, are available on Zenodo~\cite{data_repo}.

\section*{Acknowledgments}
The European Union Horizon 2020 Future and Emerging Technologies (FET) Grant Agreement No. 828966 --- QAFM and the Swedish SSF Grant No. ITM17--0343 supported this work. We thank the Quantum-Limited Atomic Force Microscopy (QAFM) team for fruitful discussions: T. Glatzel, M. Zutter, E. Tholén, D. Forchheimer, I. Ignat, M. Kwon, and D. Platz.

\section*{Conflict of Interest}
The authors have no conflicts to disclose.

\begin{appendix}
\setcounter{figure}{0}
\makeatletter 
\renewcommand{\thefigure}{A\@arabic\c@figure}
\makeatother

\section{Measurement setup}~\label{app:measurement_setup}

\begin{figure}[h]
    \centering
    \includegraphics[width=0.6\linewidth]{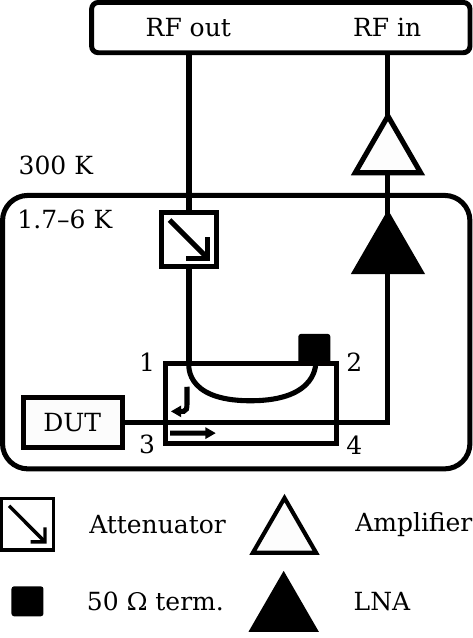}
    \caption{\label{fig:measurement_setup}
        Schematic of the cryogenic measurement setup which is inserted into a variable temperature physical properties measurement system (DynaCool, Quantum Design). The excitation signal is attenuated at low temperature and passed to the sample through a directional coupler. The reflected signal is amplified with a cryogenic low-noise amplifier (Low Noise Factory LNF-LNC4\textunderscore8E with \SI{40}{\decibel} of gain) and a room-temperature amplifier (Mini-Circuit ZX60-83LN-S+ with \SI{21}{\decibel} of gain).
    }
\end{figure}

The measurements employ a digital microwave synthesis and analysis platform operated as a vector network analyzer. The frequency response of the resonator is determined from a transmission measurement through a $\SI{20}{\decibel}$ directional coupler, as shown in Fig.~\ref{fig:measurement_setup}.

\begin{figure}[h]
    \centering
    \includegraphics[width=\linewidth]{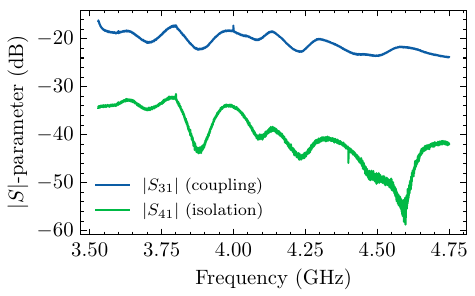}
    \caption{\label{fig:directional_coupler}
        Scattering parameters of the directional coupler. The coupling corresponds to the scattering parameter $S_{31}$ while the isolation is given by $S_{41}$. The ratio $D = S_{41}/S_{31}$ yields the directivity of the directional coupler.
    }
\end{figure}

To extract the reflection coefficient $\Gamma(\omega)$ of the resonator, we need to determine the scattering matrix of the directional coupler,
\begin{equation}
    \begin{pmatrix}
    V_{1}^{-}\\
    V_{2}^{-}\\
    V_{3}^{-}\\
    V_{4}^{-}
    \end{pmatrix}
    =
    \begin{pmatrix}
    0 & T & C & I \\
    T & 0 & I & C \\
    C & I & 0 & T \\
    I & C & T & 0 \\
    \end{pmatrix}
    \begin{pmatrix}
    V_{1}^{+}\\
    V_{2}^{+} \\
    V_{3}^{+}\\
    V_{4}^{+}
    \end{pmatrix},
    \label{eqn:directional_coupler_scattering}
\end{equation}
where $T\equiv S_{21}\approx 1$ is the transmission coefficient, $C\equiv S_{31}$ is the coupling and $I \equiv S_{41}$ is the isolation. The measured signal is the amplified voltage wave $V_{4}^{-}$ propagating out of port 4.
Given $V_{3}^{+}=\Gamma(\omega)V_{3}^{-}$ and $V_{2}^{+}=V_{4}^{+}=0$,
\begin{equation}
    V_{4}^{-}=V_{1}^{+}(I+C\Gamma(\omega)).
    \label{eqn:input_voltage}
\end{equation}
 The drive $V_{\mathrm{RFout}}$ is connected to port 1 of the directional coupler through a total attenuation $A=\SI{30}{\decibel}$ and the readout $V_{\mathrm{RFin}}$ from port 4 passes through a total gain $G=\SI{61}{\decibel}$. For $V_{1}^{+}=AV_{\mathrm{RFout}}$ and $V_{\mathrm{RFin}}=GV_{4}^{-}$,
 \begin{equation}
    V_{\mathrm{RFin}} = G A V_{\mathrm{RFout}} C ( D + \Gamma(\omega) ),
\label{eqn:input_voltage_directional_coupler}
\end{equation}
where $D=I/C$ is the directivity. Figure~\ref{fig:directional_coupler} shows the measured coupling $C \equiv S_{31}$ and isolation $I \equiv S_{41}$ as a function of frequency.

\end{appendix}

\pagebreak

\bibliographystyle{unsrtnat}

\begin{thebibliography}{35}
\providecommand{\natexlab}[1]{#1}
\providecommand{\url}[1]{\texttt{#1}}
\expandafter\ifx\csname urlstyle\endcsname\relax
  \providecommand{\doi}[1]{doi: #1}\else
  \providecommand{\doi}{doi: \begingroup \urlstyle{rm}\Url}\fi

\bibitem[Niepce et~al.(2019)Niepce, Burnett, and Bylander]{niepce2019nbn}
David Niepce, Jonathan Burnett, and Jonas Bylander.
\newblock High kinetic inductance nbn nanowire superinductors.
\newblock \emph{Phys. Rev. Applied}, 11:\penalty0 044014, Apr 2019.
\newblock \doi{10.1103/PhysRevApplied.11.044014}.
\newblock URL \url{https://link.aps.org/doi/10.1103/PhysRevApplied.11.044014}.

\bibitem[Gr{\"u}nhaupt et~al.(2019)Gr{\"u}nhaupt, Spiecker, Gusenkova, Maleeva,
  Skacel, Takmakov, Valenti, Winkel, Rotzinger, Wernsdorfer, Ustinov, and
  Pop]{grunhaupt2019granularaluminiumforquantumcircuits}
Lukas Gr{\"u}nhaupt, Martin Spiecker, Daria Gusenkova, Nataliya Maleeva,
  Sebastian~T. Skacel, Ivan Takmakov, Francesco Valenti, Patrick Winkel, Hannes
  Rotzinger, Wolfgang Wernsdorfer, Alexey~V. Ustinov, and Ioan~M. Pop.
\newblock Granular aluminium as a superconducting material for high-impedance
  quantum circuits.
\newblock \emph{Nature Materials}, 18\penalty0 (8):\penalty0 816--819, Aug
  2019.
\newblock ISSN 1476-4660.
\newblock \doi{10.1038/s41563-019-0350-3}.
\newblock URL \url{https://doi.org/10.1038/s41563-019-0350-3}.

\bibitem[Shaikhaidarov et~al.(2022)Shaikhaidarov, Kim, Dunstan, Antonov,
  Linzen, Ziegler, Golubev, Antonov, Il’ichev, and
  Astafiev]{shaikhaidarov2022quantized}
Rais~S Shaikhaidarov, Kyung~Ho Kim, Jacob~W Dunstan, Ilya~V Antonov, Sven
  Linzen, Mario Ziegler, Dmitry~S Golubev, Vladimir~N Antonov, Evgeni~V
  Il’ichev, and Oleg~V Astafiev.
\newblock Quantized current steps due to the ac coherent quantum phase-slip
  effect.
\newblock \emph{Nature}, 608\penalty0 (7921):\penalty0 45--49, 2022.

\bibitem[Rieger et~al.(2023{\natexlab{a}})Rieger, G{\"u}nzler, Spiecker,
  Paluch, Winkel, Hahn, Hohmann, Bacher, Wernsdorfer, and
  Pop]{rieger2023graluminium}
D.~Rieger, S.~G{\"u}nzler, M.~Spiecker, P.~Paluch, P.~Winkel, L.~Hahn, J.~K.
  Hohmann, A.~Bacher, W.~Wernsdorfer, and I.~M. Pop.
\newblock Granular aluminium nanojunction fluxonium qubit.
\newblock \emph{Nature Materials}, 22\penalty0 (2):\penalty0 194--199, 02
  2023{\natexlab{a}}.
\newblock ISSN 1476-4660.
\newblock \doi{10.1038/s41563-022-01417-9}.
\newblock URL \url{https://doi.org/10.1038/s41563-022-01417-9}.

\bibitem[Frunzio et~al.(2005)Frunzio, Wallraff, Schuster, Majer, and
  Schoelkopf]{frunzio2005nb}
L.~Frunzio, A.~Wallraff, D.~Schuster, J.~Majer, and R.~Schoelkopf.
\newblock Fabrication and characterization of superconducting circuit qed
  devices for quantum computation.
\newblock \emph{IEEE Transactions on Applied Superconductivity}, 15\penalty0
  (2):\penalty0 860--863, 2005.
\newblock \doi{10.1109/TASC.2005.850084}.

\bibitem[Anferov et~al.(2020)Anferov, Suleymanzade, Oriani, Simon, and
  Schuster]{anferov2020nbn}
Alexander Anferov, Aziza Suleymanzade, Andrew Oriani, Jonathan Simon, and
  David~I. Schuster.
\newblock Millimeter-wave four-wave mixing via kinetic inductance for quantum
  devices.
\newblock \emph{Phys. Rev. Appl.}, 13:\penalty0 024056, Feb 2020.
\newblock \doi{10.1103/PhysRevApplied.13.024056}.
\newblock URL \url{https://link.aps.org/doi/10.1103/PhysRevApplied.13.024056}.

\bibitem[Frasca et~al.(2023)Frasca, Arabadzhiev, de~Puechredon, Oppliger,
  Jouanny, Musio, Scigliuzzo, Minganti, Scarlino, and Charbon]{frasca2023nbn}
S.~Frasca, I.N. Arabadzhiev, S.Y.~Bros de~Puechredon, F.~Oppliger, V.~Jouanny,
  R.~Musio, M.~Scigliuzzo, F.~Minganti, P.~Scarlino, and E.~Charbon.
\newblock Nbn films with high kinetic inductance for high-quality compact
  superconducting resonators.
\newblock \emph{Phys. Rev. Appl.}, 20:\penalty0 044021, Oct 2023.
\newblock \doi{10.1103/PhysRevApplied.20.044021}.
\newblock URL \url{https://link.aps.org/doi/10.1103/PhysRevApplied.20.044021}.

\bibitem[Leduc et~al.(2010)Leduc, Bumble, Day, Eom, Gao, Golwala, Mazin,
  McHugh, Merrill, Moore, Noroozian, Turner, and Zmuidzinas]{leduc2010tin}
Henry~G. Leduc, Bruce Bumble, Peter~K. Day, Byeong~Ho Eom, Jiansong Gao, Sunil
  Golwala, Benjamin~A. Mazin, Sean McHugh, Andrew Merrill, David~C. Moore, Omid
  Noroozian, Anthony~D. Turner, and Jonas Zmuidzinas.
\newblock Titanium nitride films for ultrasensitive microresonator detectors.
\newblock \emph{Applied Physics Letters}, 97\penalty0 (10):\penalty0 102509,
  2010.
\newblock \doi{10.1063/1.3480420}.
\newblock URL \url{https://doi.org/10.1063/1.3480420}.

\bibitem[Swenson et~al.(2013)Swenson, Day, Eom, Leduc, Llombart, McKenney,
  Noroozian, and Zmuidzinas]{swenson2013tin}
L.~J. Swenson, P.~K. Day, B.~H. Eom, H.~G. Leduc, N.~Llombart, C.~M. McKenney,
  O.~Noroozian, and J.~Zmuidzinas.
\newblock Operation of a titanium nitride superconducting microresonator
  detector in the nonlinear regime.
\newblock \emph{Journal of Applied Physics}, 113\penalty0 (10):\penalty0
  104501, 2013.
\newblock \doi{10.1063/1.4794808}.
\newblock URL \url{https://doi.org/10.1063/1.4794808}.

\bibitem[Joshi et~al.(2022)Joshi, Chen, LeDuc, Day, and
  Mirhosseini]{joshi2022tin}
Chaitali Joshi, Wenyuan Chen, Henry~G. LeDuc, Peter~K. Day, and Mohammad
  Mirhosseini.
\newblock Strong kinetic-inductance kerr nonlinearity with titanium nitride
  nanowires.
\newblock \emph{Phys. Rev. Appl.}, 18:\penalty0 064088, Dec 2022.
\newblock \doi{10.1103/PhysRevApplied.18.064088}.
\newblock URL \url{https://link.aps.org/doi/10.1103/PhysRevApplied.18.064088}.

\bibitem[Tanner et~al.(2010)Tanner, Natarajan, Pottapenjara, O’Connor,
  Warburton, Hadfield, Baek, Nam, Dorenbos, Ureña, Zijlstra, Klapwijk, and
  Zwiller]{tanner2010nbtin}
M.~G. Tanner, C.~M. Natarajan, V.~K. Pottapenjara, J.~A. O’Connor, R.~J.
  Warburton, R.~H. Hadfield, B.~Baek, S.~Nam, S.~N. Dorenbos, E.~Bermúdez
  Ureña, T.~Zijlstra, T.~M. Klapwijk, and V.~Zwiller.
\newblock {Enhanced telecom wavelength single-photon detection with NbTiN
  superconducting nanowires on oxidized silicon}.
\newblock \emph{Applied Physics Letters}, 96\penalty0 (22):\penalty0 221109, 06
  2010.
\newblock ISSN 0003-6951.
\newblock \doi{10.1063/1.3428960}.
\newblock URL \url{https://doi.org/10.1063/1.3428960}.

\bibitem[Barends et~al.(2010)Barends, Vercruyssen, Endo, de~Visser, Zijlstra,
  Klapwijk, and Baselmans]{barends2010nbtin}
R.~Barends, N.~Vercruyssen, A.~Endo, P.~J. de~Visser, T.~Zijlstra, T.~M.
  Klapwijk, and J.~J.~A. Baselmans.
\newblock Reduced frequency noise in superconducting resonators.
\newblock \emph{Applied Physics Letters}, 97\penalty0 (3):\penalty0 033507,
  2010.
\newblock \doi{10.1063/1.3467052}.
\newblock URL \url{https://doi.org/10.1063/1.3467052}.

\bibitem[Ho~Eom et~al.(2012)Ho~Eom, Day, LeDuc, and Zmuidzinas]{hoeom2012nbtin}
Byeong Ho~Eom, Peter~K. Day, Henry~G. LeDuc, and Jonas Zmuidzinas.
\newblock A wideband, low-noise superconducting amplifier with high dynamic
  range.
\newblock \emph{Nature Physics}, 8\penalty0 (8):\penalty0 623--627, Aug 2012.
\newblock ISSN 1745-2481.
\newblock \doi{10.1038/nphys2356}.
\newblock URL \url{https://doi.org/10.1038/nphys2356}.

\bibitem[Samkharadze et~al.(2016)Samkharadze, Bruno, Scarlino, Zheng,
  DiVincenzo, DiCarlo, and Vandersypen]{samkharadze2016nbtin}
N.~Samkharadze, A.~Bruno, P.~Scarlino, G.~Zheng, D.~P. DiVincenzo, L.~DiCarlo,
  and L.~M.~K. Vandersypen.
\newblock High-kinetic-inductance superconducting nanowire resonators for
  circuit qed in a magnetic field.
\newblock \emph{Phys. Rev. Appl.}, 5:\penalty0 044004, Apr 2016.
\newblock \doi{10.1103/PhysRevApplied.5.044004}.
\newblock URL \url{https://link.aps.org/doi/10.1103/PhysRevApplied.5.044004}.

\bibitem[Steinhauer et~al.(2020)Steinhauer, Yang, Gyger, Lettner,
  Errando-Herranz, Jöns, Baghban, Gallo, Zichi, and
  Zwiller]{steinhauer2020nbtin}
Stephan Steinhauer, Lily Yang, Samuel Gyger, Thomas Lettner, Carlos
  Errando-Herranz, Klaus~D. Jöns, Mohammad~Amin Baghban, Katia Gallo, Julien
  Zichi, and Val Zwiller.
\newblock Nbtin thin films for superconducting photon detectors on photonic and
  two-dimensional materials.
\newblock \emph{Applied Physics Letters}, 116\penalty0 (17):\penalty0 171101,
  2020.
\newblock \doi{10.1063/1.5143986}.
\newblock URL \url{https://doi.org/10.1063/1.5143986}.

\bibitem[Burdastyh et~al.(2020)Burdastyh, Postolova, Proslier, Ustavshikov,
  Antonov, Vinokur, and Mironov]{burdastyh2020nbtin}
M.~V. Burdastyh, S.~V. Postolova, T.~Proslier, S.~S. Ustavshikov, A.~V.
  Antonov, V.~M. Vinokur, and A.~Yu. Mironov.
\newblock Superconducting phase transitions in disordered nbtin films.
\newblock \emph{Scientific Reports}, 10\penalty0 (1):\penalty0 1471, Jan 2020.
\newblock ISSN 2045-2322.
\newblock \doi{10.1038/s41598-020-58192-3}.
\newblock URL \url{https://doi.org/10.1038/s41598-020-58192-3}.

\bibitem[Bretz-Sullivan et~al.(2022)Bretz-Sullivan, Lewis, Lima-Sharma, Lidsky,
  Smyth, Harris, Venuti, Eley, and Lu]{bretz-sullivan2022nbtin}
Terence~M. Bretz-Sullivan, Rupert~M. Lewis, Ana~L. Lima-Sharma, David Lidsky,
  Christopher~M. Smyth, C.~Thomas Harris, Michael Venuti, Serena Eley, and
  Tzu-Ming Lu.
\newblock {High kinetic inductance NbTiN superconducting transmission line
  resonators in the very thin film limit}.
\newblock \emph{Applied Physics Letters}, 121\penalty0 (5):\penalty0 052602, 08
  2022.
\newblock ISSN 0003-6951.
\newblock \doi{10.1063/5.0100961}.
\newblock URL \url{https://doi.org/10.1063/5.0100961}.

\bibitem[Parker et~al.(2022)Parker, Savytskyi, Vine, Laucht, Duty, Morello,
  Grimsmo, and Pla]{parker2022nbtin}
Daniel~J. Parker, Mykhailo Savytskyi, Wyatt Vine, Arne Laucht, Timothy Duty,
  Andrea Morello, Arne~L. Grimsmo, and Jarryd~J. Pla.
\newblock Degenerate parametric amplification via three-wave mixing using
  kinetic inductance.
\newblock \emph{Phys. Rev. Appl.}, 17:\penalty0 034064, Mar 2022.
\newblock \doi{10.1103/PhysRevApplied.17.034064}.
\newblock URL \url{https://link.aps.org/doi/10.1103/PhysRevApplied.17.034064}.

\bibitem[Korzh et~al.(2020)Korzh, Zhao, Allmaras, Frasca, Autry, Bersin, Beyer,
  Briggs, Bumble, Colangelo, Crouch, Dane, Gerrits, Lita, Marsili, Moody,
  Pe{\~{n}}a, Ramirez, Rezac, Sinclair, Stevens, Velasco, Verma, Wollman, Xie,
  Zhu, Hale, Spiropulu, Silverman, Mirin, Nam, Kozorezov, Shaw, and
  Berggren]{korzh2020sspd}
Boris Korzh, Qing-Yuan Zhao, Jason~P. Allmaras, Simone Frasca, Travis~M. Autry,
  Eric~A. Bersin, Andrew~D. Beyer, Ryan~M. Briggs, Bruce Bumble, Marco
  Colangelo, Garrison~M. Crouch, Andrew~E. Dane, Thomas Gerrits, Adriana~E.
  Lita, Francesco Marsili, Galan Moody, Cristi{\'a}n Pe{\~{n}}a, Edward
  Ramirez, Jake~D. Rezac, Neil Sinclair, Martin~J. Stevens, Angel~E. Velasco,
  Varun~B. Verma, Emma~E. Wollman, Si~Xie, Di~Zhu, Paul~D. Hale, Maria
  Spiropulu, Kevin~L. Silverman, Richard~P. Mirin, Sae~Woo Nam, Alexander~G.
  Kozorezov, Matthew~D. Shaw, and Karl~K. Berggren.
\newblock Demonstration of sub-3 ps temporal resolution with a superconducting
  nanowire single-photon detector.
\newblock \emph{Nature Photonics}, 14\penalty0 (4):\penalty0 250--255, Apr
  2020.
\newblock ISSN 1749-4893.
\newblock \doi{10.1038/s41566-020-0589-x}.
\newblock URL \url{https://doi.org/10.1038/s41566-020-0589-x}.

\bibitem[Day et~al.(2003)Day, LeDuc, Mazin, Vayonakis, and
  Zmuidzinas]{day2003mkid}
Peter~K. Day, Henry~G. LeDuc, Benjamin~A. Mazin, Anastasios Vayonakis, and
  Jonas Zmuidzinas.
\newblock A broadband superconducting detector suitable for use in large
  arrays.
\newblock \emph{Nature}, 425\penalty0 (6960):\penalty0 817--821, Oct 2003.
\newblock ISSN 1476-4687.
\newblock \doi{10.1038/nature02037}.
\newblock URL \url{https://doi.org/10.1038/nature02037}.

\bibitem[Zmuidzinas(2012)]{zmuidzinas2012review}
Jonas Zmuidzinas.
\newblock Superconducting microresonators: Physics and applications.
\newblock \emph{Annual Review of Condensed Matter Physics}, 3\penalty0
  (1):\penalty0 169--214, 2012.
\newblock \doi{10.1146/annurev-conmatphys-020911-125022}.
\newblock URL \url{https://doi.org/10.1146/annurev-conmatphys-020911-125022}.

\bibitem[Roos et~al.(2023)Roos, Scarano, Arvidsson, Holmgren, and
  Haviland]{roos2023kimec}
August~K. Roos, Ermes Scarano, Elisabet~K. Arvidsson, Erik Holmgren, and
  David~B. Haviland.
\newblock Kinetic inductive electromechanical transduction for nanoscale force
  sensing.
\newblock \emph{Phys. Rev. Appl.}, 20:\penalty0 024022, 08 2023.
\newblock \doi{10.1103/PhysRevApplied.20.024022}.
\newblock URL \url{https://link.aps.org/doi/10.1103/PhysRevApplied.20.024022}.

\bibitem[Rammer(1988)]{rammer1988magnetic}
J~Rammer.
\newblock Magnetic penetration depth in yba2cu3o9-$\delta$: Evidence for strong
  electron-phonon coupling?
\newblock \emph{Europhysics Letters}, 5\penalty0 (1):\penalty0 77, 1988.

\bibitem[Bardeen et~al.(1957)Bardeen, Cooper, and
  Schrieffer]{bardeen1957theory}
J.~Bardeen, L.~N. Cooper, and J.~R. Schrieffer.
\newblock Theory of superconductivity.
\newblock \emph{Phys. Rev.}, 108:\penalty0 1175--1204, Dec 1957.
\newblock \doi{10.1103/PhysRev.108.1175}.
\newblock URL \url{https://link.aps.org/doi/10.1103/PhysRev.108.1175}.

\bibitem[Roos et~al.(2024)Roos, Scarano, Arvidsson, Holmgren, and
  Haviland]{roos2024design}
August~K. Roos, Ermes Scarano, Elisabet~K. Arvidsson, Erik Holmgren, and
  David~B. Haviland.
\newblock Design, fabrication, and characterization of kinetic-inductive force
  sensors for scanning probe applications.
\newblock \emph{Beilstein Journal of Nanotechnology}, 15:\penalty0 242--255,
  2024.
\newblock ISSN 2190-4286.
\newblock \doi{10.3762/bjnano.15.23}.
\newblock URL \url{https://doi.org/10.3762/bjnano.15.23}.

\bibitem[Khan et~al.(2022)Khan, Khudchenko, Chekushkin, and
  Koshelets]{khan2022characterization}
Fedor Khan, AV~Khudchenko, AM~Chekushkin, and Valery~P Koshelets.
\newblock Characterization of the parameters of superconducting nbn and nbtin
  films using parallel plate resonator.
\newblock \emph{IEEE Transactions on Applied Superconductivity}, 32\penalty0
  (4):\penalty0 1--5, 2022.

\bibitem[Watanabe et~al.(1994)Watanabe, Yoshida, and
  Kohjiro]{watanabe1994kinetic}
Koki Watanabe, Keiji Yoshida, and Takeshi~Aoki Kohjiro.
\newblock Kinetic inductance of superconducting coplanar waveguides.
\newblock \emph{Japanese journal of applied physics}, 33\penalty0
  (10R):\penalty0 5708, 1994.

\bibitem[Pozar(2011)]{pozar2011microwave}
David~M Pozar.
\newblock \emph{Microwave engineering}.
\newblock John wiley \& sons, 2011.

\bibitem[Rieger et~al.(2023{\natexlab{b}})Rieger, G\"unzler, Spiecker,
  Nambisan, Wernsdorfer, and Pop]{rieger2023microwaveresonatorfano}
D.~Rieger, S.~G\"unzler, M.~Spiecker, A.~Nambisan, W.~Wernsdorfer, and I.M.
  Pop.
\newblock Fano interference in microwave resonator measurements.
\newblock \emph{Phys. Rev. Appl.}, 20:\penalty0 014059, Jul 2023{\natexlab{b}}.
\newblock \doi{10.1103/PhysRevApplied.20.014059}.
\newblock URL \url{https://link.aps.org/doi/10.1103/PhysRevApplied.20.014059}.

\bibitem[Tinkham(2004)]{tinkham2004introduction}
Michael Tinkham.
\newblock \emph{Introduction to superconductivity}.
\newblock Courier Corporation, 2004.

\bibitem[Annunziata et~al.(2010)Annunziata, Santavicca, Frunzio, Catelani,
  Rooks, Frydman, and Prober]{annunziata2010nanoinductors}
Anthony~J Annunziata, Daniel~F Santavicca, Luigi Frunzio, Gianluigi Catelani,
  Michael~J Rooks, Aviad Frydman, and Daniel~E Prober.
\newblock Tunable superconducting nanoinductors.
\newblock \emph{Nanotechnology}, 21\penalty0 (44):\penalty0 445202, oct 2010.
\newblock \doi{10.1088/0957-4484/21/44/445202}.
\newblock URL \url{https://dx.doi.org/10.1088/0957-4484/21/44/445202}.

\bibitem[Sheahen(1966)]{sheahen1966rulesforenergygap}
Thomas~P. Sheahen.
\newblock Rules for the energy gap and critical field of superconductors.
\newblock \emph{Phys. Rev.}, 149:\penalty0 368--370, Sep 1966.
\newblock \doi{10.1103/PhysRev.149.368}.
\newblock URL \url{https://link.aps.org/doi/10.1103/PhysRev.149.368}.

\bibitem[Linden et~al.(1994)Linden, Orlando, and Lyons]{linden1994modified}
Derek~S Linden, Terry~P Orlando, and W~Gregory Lyons.
\newblock Modified two-fluid model for superconductor surface impedance
  calculation.
\newblock \emph{IEEE Transactions on Applied Superconductivity}, 4\penalty0
  (3):\penalty0 136--142, 1994.

\bibitem[Gerhold(1998)]{gerhold1998properties}
J.~Gerhold.
\newblock Properties of cryogenic insulants.
\newblock \emph{Cryogenics}, 38\penalty0 (11):\penalty0 1063--1081, 1998.
\newblock ISSN 0011-2275.
\newblock \doi{https://doi.org/10.1016/S0011-2275(98)00094-0}.
\newblock URL
  \url{https://www.sciencedirect.com/science/article/pii/S0011227598000940}.

\bibitem[Scarano et~al.(2024)Scarano, Arvidsson, Roos, Holmgren, and
  Haviland]{data_repo}
Ermes Scarano, Elisabet~K. Arvidsson, August~K. Roos, Erik Holmgren, and
  David~B. Haviland.
\newblock Data and code for figures: Temperature dependence of microwave losses
  in lumped-element resonators made from superconducting nanowires with high
  kinetic inductance, 2024.
\newblock URL \url{https://zenodo.org/records/11656762}.

\end{thebibliography}

\end{document}